\documentclass[epjCONF,columns]{svjour}
\usepackage{graphicx}
\usepackage[varg]{txfonts} 
\usepackage[latin1]{inputenc}
\session-title{Eurasia Pacific Summer School and Conference on Strongly Correlated Electrons}
\begin{document}
\title{Superconducting fluctuations and pseudogap in high-$T_c$ cuprates}
\author{F. Rullier-Albenque\inst{1}\fnmsep\thanks{\email{florence.albenque-rullier@cea.fr}} \and H. Alloul\inst{2}}
\institute{Service de Physique de l'Etat Condens\'e, Orme des Merisiers, CEA Saclay
(CNRS URA 2464), 91191 Gif sur Yvette cedex, France \and Laboratoire de Physique des Solides, UMR CNRS 8502,
Universit\'e Paris Sud, 91405 Orsay, France}
\abstract{
Large pulsed magnetic fields up to 60 Tesla are used to suppress the contribution of superconducting
fluctuations (SCF) to the ab-plane conductivity above $T_c$ in a series of YBa$_2$Cu$_3$O$_{6+x}$. 
These experiments allow us to determine the
field $H_c^{\prime}(T)$ and the temperature $T_c^{\prime}$ above which the SCFs are fully suppressed. 
A careful investigation near optimal doping shows that $T_c^{\prime}$ is higher than the pseudogap
temperature $T^{\star}$, which is an unambiguous evidence that the pseudogap cannot be assigned to 
preformed pairs. 
Accurate determinations of the SCF contribution to the conductivity versus temperature and magnetic 
field have been achieved. They can be accounted for by thermal fluctuations following the Ginzburg-Landau
scheme for nearly optimally doped samples. A phase fluctuation contribution might be invoked for the 
most underdoped samples in a $T$ range which increases when controlled disorder is introduced by 
electron irradiation. Quantitative analysis of the fluctuating magnetoconductance allows us to determine 
the critical field $H_{c2}(0)$ which is found to be be quite similar to $H_c^{\prime}(0)$ and to 
increase with hole doping.
Studies of the incidence of disorder on both $T_c^{\prime}$ and $T^{\star}$ allow us to to propose a 
three dimensional phase diagram including a disorder axis, which allows to explain most observations 
done in other cuprate families.
} 
\maketitle
\section{Introduction}
\label{intro}
\begin{figure}
\centerline{\includegraphics[width=0.8\columnwidth]{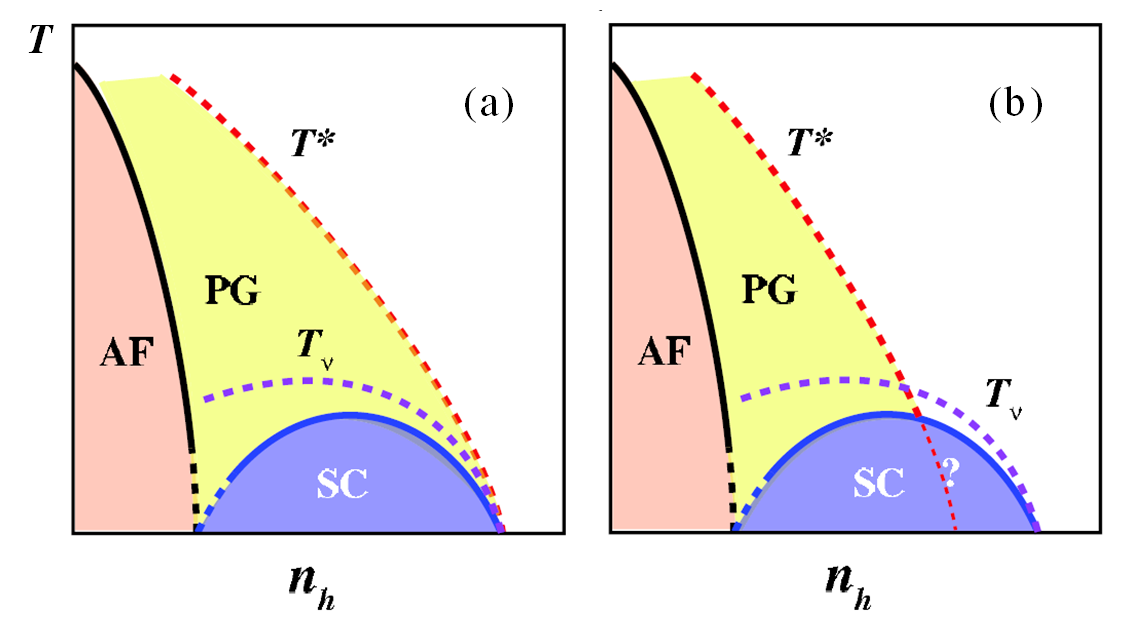}}
\caption{Different scenarios for the phase diagram of the high-$T_c$ cuprates. While in (a) $T^*$ merges 
with $T_c$ in the strongly overdoped regime, in (b) $T^{\star}$ intersects the superconducting dome 
near optimal doping. $T_{\nu}$ represents the onset of the Nernst signal.}
\label{fig:phase-diagram}      
\end{figure}

One of the most puzzling feature of the high-$T_c$ cuprates is the existence of the so-called pseudogap phase in the underdoped region of their phase diagram. After the first evidence of an anomalous drop of the spin suusceptibility detected by NMR experiments in underdoped YBCO well above $T_c$ \cite{Alloul-1989}, a lot of unusual properties have been observed in the pseudogap phase \cite{Timusk}. Quite surprisingly, despite the huge effort to characterize this phase in the different cuprate families, there does not exist up to now a unique representation of the pseudogap line $T^{\star}$ as illustrated in Fig.\ref{fig:phase-diagram}. Either $T^*$ is found to merge with the superconducting dome in the overdoped part of the phase diagram, or to cross it near optimal doping. These different representations are associated with different lines of thought. In the first case, it has been proposed that the pseudogap could be ascribed to the formation of superconducting pairs with strong phase fluctuations \cite{Emery}. This scenario has been supported by the observation of a large Nernst effect and of diamagnetism above $T_c$, which delineates another line $T_{\nu}$ below which strong superconducting fluctuations and/or vortices persist in the normal state \cite{Wang-PRB2006}. In the second approach, the pseudogap and the superconducting phases arise from different, even competing, underlying mechanisms and are associated to different energy scales \cite{Hufner}.

In the following, we will present our results on superconducting fluctuations for a series of
YBa$_2$Cu$_3$O$_{6+x}$ single crystals from underdoping to slightly overdoping. We have used an original
method based on the measurements of the magnetoresistance in high pulsed magnetic fields 
\cite{FRA-HF} \cite{FRA-PRB2011}. This allows us to determine the normal state resistivity and to 
extract with high accuracy the superconducting
fluctuations (SCF) contributions to the conductivity and their dependences as a function of temperature 
and magnetic field. We are thus able to determine the threshold values of the magnetic field 
$H_{c}^{\prime}$ and temperature $T_{c}^{\prime}$ above which the normal state is completely restored. 

In the same set of transport data, we can compare here the values of $T_{c}^{\prime}$ and of the 
pseudogap temperature $T^{\star}$ as a function of doping \cite{Alloul-EPL2010}. We will show how 
our results can be analysed in the framework of the Ginzburg-Landau model, making it possible to 
extract microscopic parameters of the superconducting state such as the zero-temperature coherence length. 
The effect of disorder introduced by electron irradiation at low temperature will be also presented. 

\section{Experimental}
\label{sec:exp}
Details on the experimental conditions concerning the different single crystals and the high-field
experiments as well as the method used to extract the SCF contribution to the conductivity are given 
in ref.\cite{FRA-PRB2011}. Four different single crystals of YBa$_2$Cu$_3$O$_{6+x}$ have been studied. 
They are labelled with respect to their critical temperatures measured at the mid-point of the resistive
transition: two underdoped samples UD57 and UD85, an optimally doped sample OPT93.6 and a slightly 
overdoped one OD92.5, corresponding to oxygen contents of approximately 6.54, 6.8, 6.91 and 6.95
respectively. Some of these samples have been irradiated by electrons at low $T$, which allows us 
to introduce a well controlled concentration of defects in the CuO$_2$ planes \cite{Legris}.

\begin{figure}
\centerline{\includegraphics[width=0.8\columnwidth]{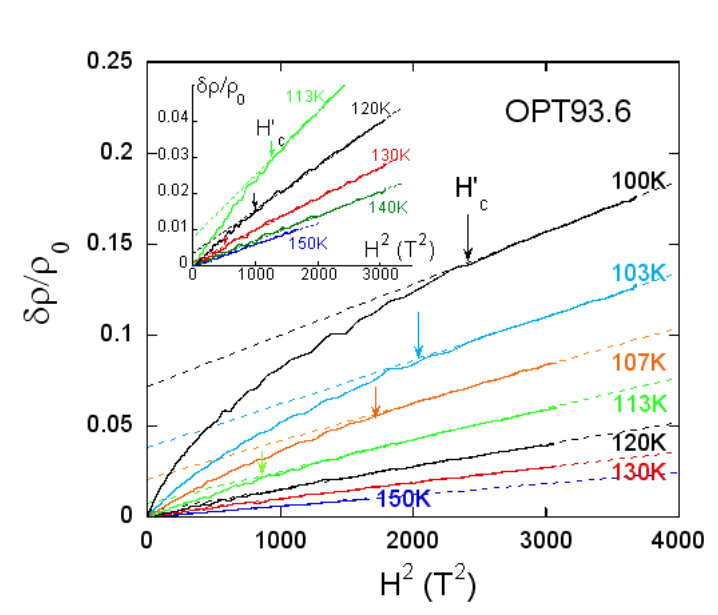}}
\caption{Field variation of the resistivity increase normalized to its zero-field value $\delta\rho/\rho_0$ plotted versus $H^{2}$ for decreasing temperatures down to $T \backsimeq T_c$ in the optimally doped sample OPT93.6. The inset shows an enlargement of the curves for the highest temperatures.}
\label{fig:MR-OPT}       
\end{figure}

The transverse MR of the different samples have been measured in a pulse field magnet up to 60T at 
the LNCMI in Toulouse. An example of the transverse MR curves measured on the OPT93.6 sample is illustrated
in fig.\ref{fig:MR-OPT} for $T$ ranging from above $T_c$ to 150K. In the normal state well above $T_c$, it 
is well known that the transverse MR increases as $H^2$. This is indeed what is found also here for $H$ up 
to 60T and for $T\gtrsim140$K (see inset of fig.\ref{fig:MR-OPT}. At lower $T$, some downward departure 
from this $H^2$ behavior is observed for low values of $H$ which we attribute to the destruction of SCFs by 
the magnetic field. The normal state behavior is only restored above a threshold field $H_{c}^{\prime}(T)$
which increases with decreasing temperatures. 

This experimental approach allows us to single out the 
normal state properties and determine the SCF contributions to the transport. In particular, the
extrapolation down to $H=0$ of the $H^2$ normal state MR above $H_{c}^{\prime}(T)$ gives us the value of 
the normal state resistivity $\rho_{n}(T)$. The way to extract the fluctuating conductivity and its
dependence with temperature $\Delta\sigma_{SF}(T,0)$ and magnetic field $\Delta\sigma_{SF}(T,H)$ is 
explained in details in ref.\cite{FRA-PRB2011}.

\section{Pseudogap and onset of superconducting fluctuations}
\label{sec:PG and onset}
\begin{figure}
\centerline{\includegraphics[width=0.8\columnwidth]{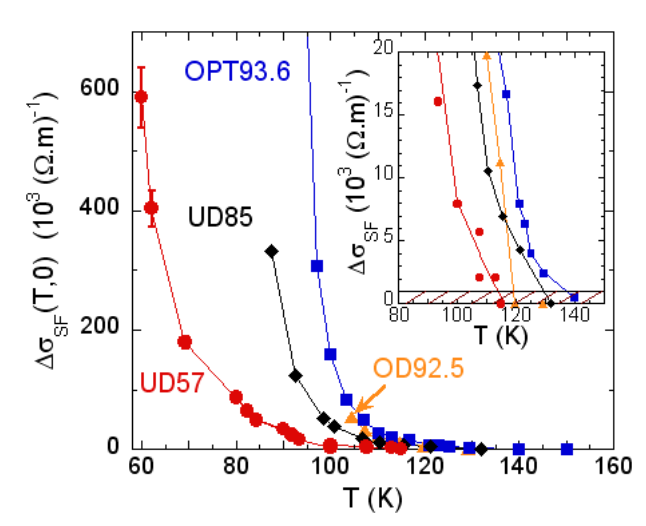}}
\caption{SCF contribution to the zero-field conductivity $\Delta\sigma_{SF}(T,0)$ for the four YBCO 
samples studied here. The enlargement of the high-$T$ range shown in the inset visualizes the criterion 
used to define the onset temperature $T_{}^{\prime}$ for SCFs. Lines are guides for the eyes.}
\label{fig:dsigma(T)}      
\end{figure}

The $T$ dependences of the zero-field superconducting conductivities $\Delta\sigma_{SF}(T,0)$ are reported 
in Fig.\ref{fig:dsigma(T)} for the four samples considered here. Let us note that this quantity 
vanishes very fast, allowing us to define very precisely an onset temperature $T_{c}^{\prime}$ for SCFs. 
As indicated in the inset of the figure, $T_{}^{\prime}$ is defined as the temperature above 
which $\Delta\sigma_{SF}(T,0)$ is lower than 1 x $10^{3}(\Omega m)^{-1}$. We observe that $T_{c}^{\prime}$ 
is always found larger that the onset of Nernst signal measured on the 
same samples \cite{FRA-Nernst}. 
One can point out here that
$T_{c}^{\prime}$ is only slightly dependent on hole doping, increasing from $\backsim120$K to 
$\backsim140$k from the UD57 sample to the optimally doped one OPT92.6. This is very similar to what 
has been found from Nernst or magnetization experiments in Bi2212 \cite{Wang-PRL2}. However this strongly
contrasts with the pseudogap temperature $T^{\star}$ which decreases with increasing doping. 

In transport measurements, the pseudogap temperature is usually determined from the downward departure of
$\rho_{ab}$ from its linear high-$T$ variation. In the case of the underdoped samples UD57 and UD85, 
this criterion yields $T^{\star}$ equal to $\backsim300$K and $\backsim210$K respectively, well 
above $T_{c}^{\prime}$. 
\begin{figure}[hb]
\centerline{\includegraphics[width=0.8\columnwidth]{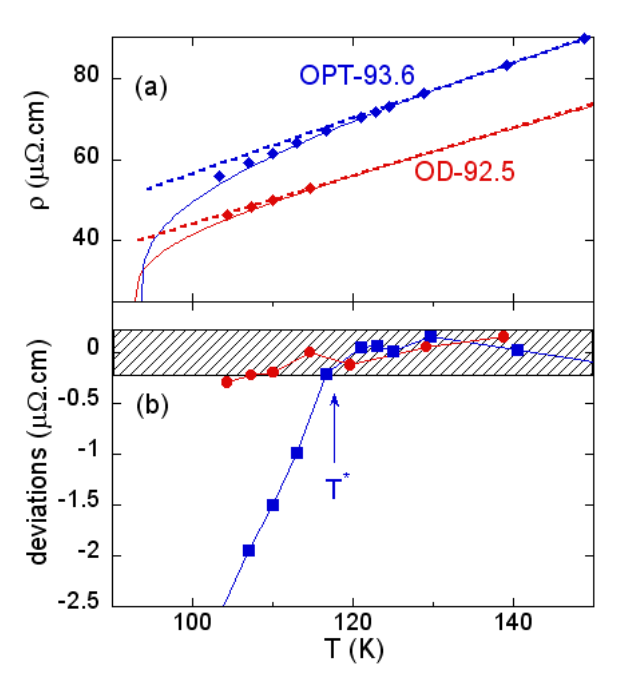}}
\caption{(a) T variations of the zero-field resistivities and $\rho_{n}(T,0)$ extracted from high-field data for OPT93.6 and OD92.5 samples. (b) deviations from linearity of $\rho_{n}(T,0)$ allowing to determine the pseudogap $T^{\star}$ for the OPT93.6 sample.}
\label{fig:rho_OPT-OD}      
\end{figure}

\begin{figure}
\centerline{\includegraphics[width=0.8\columnwidth]{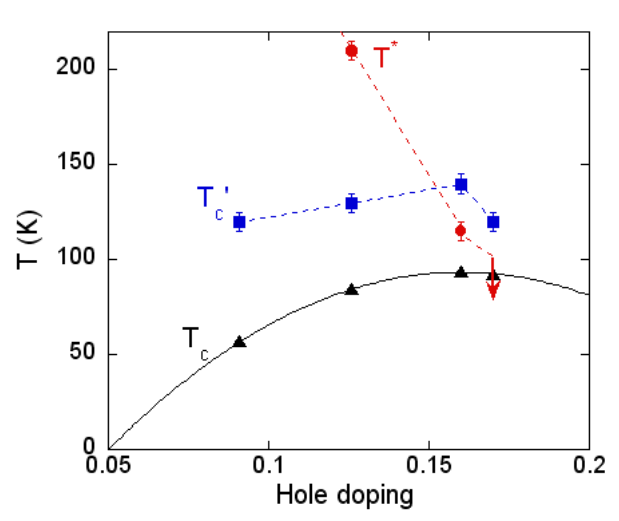}}
\caption{The values of $T_{c}^{\prime}$ (squares) and $T^{\star}$ (circles) are plotted versus the hole doping for the four samples studied. The solid line indicates the superconducting dome. Contrary to $T_{c}^{\prime}$ that is rather insensitive to hole doping, $T^{\star}$ is found to decrease with increasing doping and intersects the $T_{c}^{\prime}$ line near optimal doping \cite{Alloul-EPL2010}.}
\label{fig:onset-PG}      
\end{figure}

The situation is more delicate for the optimally doped sample where 
$T^{\star}$ becomes comparable to $T_{c}^{\prime}$. If one considers the zero-field transport 
data reported in fig.\ref{fig:rho_OPT-OD}-a, downward departures from the linear $T$ variation are 
clearly observable but cannot be attributed straightforwardly to the SCFs or to the pseudogap. 
However, when the contribution of SCFs are suppressed by the field, the data for $\rho_{n}(T,0)$ 
still displays a downward curvature for the OPT93.6 sample, which can be now only assigned to the 
pseudogap. Consequently, we are able here to determine both the onset of SCFs and the pseudogap 
temperature within the same experimental sensitivity (see fig.\ref{fig:rho_OPT-OD}-b) 
\cite{Alloul-EPL2010}. The variations of $T_{c}^{\prime}$ and $T^{\star}$ are reported in 
fig.\ref{fig:onset-PG} versus hole doping. The fact that the $T_{c}^{\prime}$ line crosses 
the pseudogap line near optimal doping unambiguously proves that the pseudogap phase cannot 
be a precursor state for superconductivity.

\section{Quantitative analysis of the paraconductivity in the Ginzburg-Landau approach}
\label{sec:Quant_analysis}
There have been a lot of studies of the $T$ dependence of the fluctuating conductivity 
$\Delta\sigma_{SF}(T)$ in optimally doped cuprates by the past. Usually, this quantity is determined 
by assuming a linear $T$ decrease of 
the normal state resistivity down to low temperature. The results reported in fig.\ref{fig:rho_OPT-OD}(a)
clearly show that this is not the case, due to the opening of the pseudogap below $T_{c}^{\prime}$. This 
puts into question the reliability of the determinations done in many cases. 

For all our samples, except the most underdoped one UD57, our results can be well accounted for by 
gaussian fluctuations within the Ginzburg-Landau (GL) theory \cite{Larkin-Varlamov}. In this approach 
the excess fluctuating 
conductivity, called here paraconductivity, is related to the temperature dependence of $\xi(T)$, 
the superconducting correlation length of the short-lived Cooper pairs, which is expected to diverge 
with decreasing temperature as:
\begin{equation}
\xi(T)=\xi(0)/\sqrt{\epsilon}
\label{xi(T)} 
\end{equation}
where $\xi(0)$ is the zero-temperature coherence length and $\epsilon=\ln (T/T_c) \backsimeq (T-T_c)/T_c$ 
for $T \gtrsim T_c$. More generally, the temperature dependence of $\Delta\sigma_{SF}(T)$ is given by 
the Lawrence-Doniach (LD) expression which takes into account the layered structure of the high-$T_c$
cuprates \cite{Lawrence-Doniach}:
\begin{equation}
\Delta\sigma^{LD}(T)=\frac{e^{2}}{16\hslash s} \frac{1}{\epsilon \sqrt{1+2\alpha}}
\label{Eq.LD}
\end{equation}
where the coupling parameter $\alpha = 2 (\xi_{c}(T)/s)^{2}$ with $\xi_{c}(T)=\xi_{c}(0)/ \sqrt{\epsilon}$.
Sufficiently far from $T_{c}$, one expects $\xi_{c}(T) \ll s$ and Eq.\ref{Eq.LD} reduces to the 
well-known 2D Aslamazov-Larkin expression: 
\begin{equation}
\Delta\sigma^{AL}(T)=\frac{e^{2}}{16\hslash s} \epsilon^{-1}=\frac{e^{2}}{16\hslash s}\frac{\xi^{2}(T)}{\xi^{2}(0)}
\label{Eq.AL}
\end{equation}
The only parameters in this expression are the value of the interlayer distance $s$ and the 
value taken for $T_c$ which can have a huge incidence on the shape of the curve especially 
for $(T-T_{c})/T_{c}<0.01$. 
\begin{figure}
\centerline{\includegraphics[width=0.8\columnwidth]{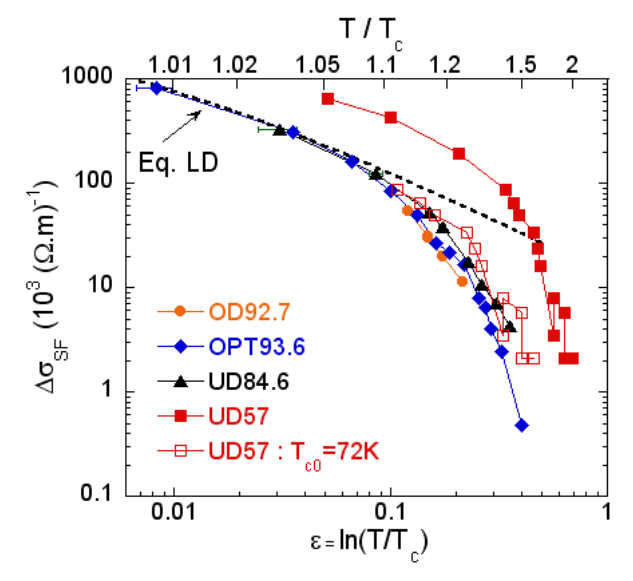}}
\caption{Superconducting fluctuation conductivity $\Delta\sigma_{SF}$ for the four pure samples 
studied plotted versus $\epsilon=ln(T/T_{c})$. Values of $T_{c}$ have been taken here at the 
midpoint of the resistive transition, and error bars for $\epsilon$ using the onset and offset values 
of $T_{c}$ are indicated. The dashed line represents the expression of Eq.(\ref{Eq.LD})
with $s=11.7$\AA. and $\xi_{c}(0) \simeq 0.9$\AA . The data for the most underdoped sample can be matched with the other ones if one takes $T_{c0}=72$K for the actual $T_c$ instead of 57.1K. Full lines are guides to the eye.} 
\label{fig:LD}     
\end{figure}

The variations of $\Delta\sigma_{SF}(T)$ are reported versus $\epsilon$ in fig.\ref{fig:LD} for the four 
hole  dopings studied. Except for the UD57 sample, it is striking to see that our experimental data 
collapse on a single curve and can be reasonably fitted by the LD expression (Eq.\ref{Eq.LD}) in the small 
temperature range $0.03 \leq \epsilon \leq 0.1$ if one takes $\xi_{c}(0) \simeq 0.9$\AA . We have 
assumed here, as usually done, that the CuO$_{2}$ bilayer constitutes the basic two-dimensional unit, 
and $s$ is then taken as the unit-cell size in the $c$ direction: $s=11.7$\AA . Based on these results, we
have also analysed Nernst data previously measured in optimally doped YBCO crystals \cite{FRA-Nernst} 
along the same lines. In the GL approach, a simple relationship can be written between $\Delta\sigma_{SF}(T)$
and the off-diagonal Peltier term $\alpha_{xy}$ as \cite{Ussishkin}:
\begin{equation}
\frac{\alpha_{xy}}{B}=\frac{8k_{B}}{3\pi \hslash}\xi(0)^{2} \Delta\sigma_{SF}(T)
\label{Eq.Nernst-sigma}
\end{equation} 
A linear dependence can be indeed verified near $T_c$ between $\Delta\sigma_{SF}$ and $\alpha_{xy}$ 
whose  slope results in a value of the zero temperature coherence length $\xi(0) \simeq 1.4$nm. 
Using $H_{c2}(0)=\Phi_0/2\pi\xi(0)^{2}$, this would lead to $H_{c2}(0) \simeq 160$T, a value very close 
to the result found below from the analysis of the magnetoconductivity. Consequently, one can 
conclude that, in optimally doped YBCO, the SCF contribution to the conductivity and the Nernst effect 
above $T_c$ can be interpreted in terms of Gaussian fluctuations only.

For the UD57 sample, $\Delta\sigma_{SF}(T)$ is found to be about a factor four larger than for the 
other dopings. Quite surprisingly, it is possible to get a good matching with the data found for the 
other dopings by assuming a effective value $T_{c0}$ different from the actual $T_c$. This is illustrated 
by the empty symbols in fig.\ref{fig:LD} using $T_{c0}=72$K. This points to an additional origin of 
SCFs below $T_{c0}$ which might be ascribed to phase fluctuations of the order parameter. This point 
is discussed in more details in ref.\cite{FRA-PRB2011}.

For all the samples, one can see in fig.\ref{fig:LD} that $\Delta\sigma_{SF}(T)$ vanishes very rapidly 
for $\epsilon \gtrsim 0.1$. This behaviour which has been noticed previously in many studies is 
particularly well defined here given the method used to extract the fluctuating conductivity. 
The cutoff which must be invoked to explain that behaviour implies that the density of 
fluctuating pairs vanishes at $T_{c}^{\prime}$.

\section{Field variation of the SCF conductivity: Onset field $H_{c}^{\prime}$ and upper critical 
field $H_{c2}$}
\label{sec:Fields}
\begin{figure}[b]
\centerline{\includegraphics[width=0.8\columnwidth]{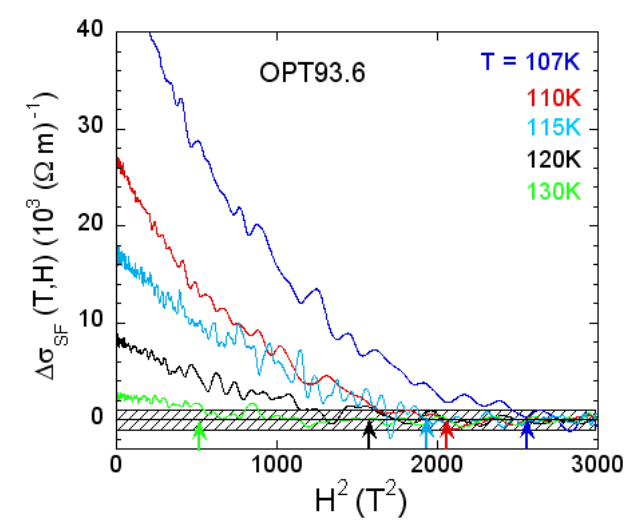}}
\caption{SC fluctuation contribution to the conductivity $\Delta\sigma_{SF}(T,H)$
in OPT93.6 plotted versus $H^{2}$. The arrows indicate the
threshold fields $H^{\prime}_{c}(T)$ taken at $\Delta\sigma_{SF}(T,H)=$1 x $10^3(\Omega.m)^{-1}$. 
The curves are plotted for increasing temperatures from top to bottom.}
\label{fig:sigma-H}      
\end{figure}

From the data reported in Fig.\ref{fig:MR-OPT}, we can extract the variation of $\Delta\sigma_{SF}(T,H)$ with the applied magnetic field and analyse how the excess conductivity is destroyed by the field. This is reported in fig.\ref{fig:sigma-H} for the optimally doped samples for $T$ between 107 and 130K.

Using the same criterion as defined above, we can thus determine the fields $H_{c}^{^\prime}(T)$ 
above which the signal is lower than 1 x $10^{3} (\Omega m)^{-1}$. As $T$ decreases, it becomes 
difficult to ascertain that the normal state is fully reached when $H_{c}^{^\prime}(T)$ becomes 
comparable to the highest available field. This makes it difficult to precisely deduce values 
of $H_{c}^{^\prime}(T)$ larger than 45T.

\begin{figure}
\centerline{\includegraphics[width=0.8\columnwidth]{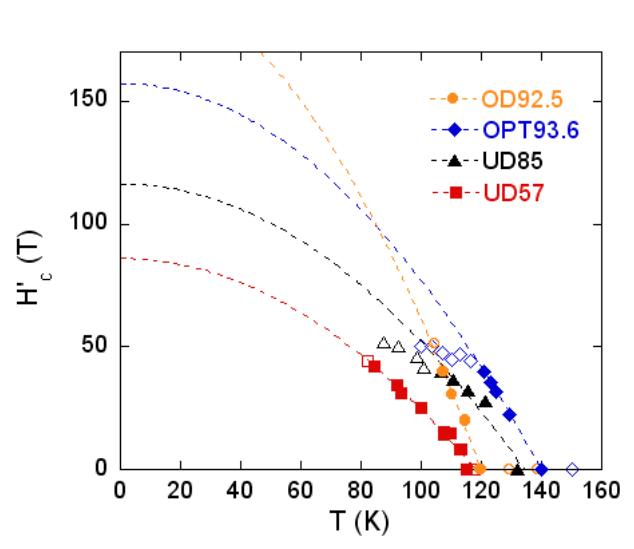}}
\caption{(color on line) The field $H_{c}^{\prime}$, at which the SC
fluctuations disappear and the normal state is fully restored, is plotted versus 
$T$ for the four pure samples studied. Dashed lines represent the fitting curves 
to Eq.\ref{Eq.H'c-T} using data with closed symbols. When $H_{c}^{\prime}(T)\gtrsim 40T$ (empty symbols), 
the data are somewhat underestimated  as the  maximum applied field is not sufficient to restore the normal state.}
\label{Fig:Hcprime}
\end{figure}

The evolution of $H_{c}^{^\prime}(T)$ are plotted in fig.\ref{Fig:Hcprime} for the four samples. 
One can see that $H_{c}^{^\prime}(T)$ drops rapidly with increasing $T$ and displays a linear 
variation near $T_{c}^{\prime}$. We have thus fitted the data using a parabolic $T$ variation 
as applied for the critical field of classical superconductors:
\begin{equation}
H_{c}^{\prime }(T)=H_{c}^{\prime }(0)[1-(T/T^{\prime }_{c})^{2}]
\label{Eq.H'c-T}
\end{equation}
The fitting curves displayed as dashed lines in Fig.\ref{Fig:Hcprime} give 
correspondingly an indication of the field $H_{c}^{\prime}(0)$
required to completely suppress the SC fluctuation contribution down to 0K. It is clear that 
$H_{c}^{\prime}(0)$ increases with hole doping and reaches a value as high as $\sim 150$ Tesla 
at optimal doping.

A precise analysis of the fluctuation magnetoconductivity $\Delta\sigma_H(T,H)$ 
is a valuable tool to extract different microscopic parameters of high-$T_{c}$ cuprates, such as 
the value of $H_{c2}(0)$ not directly accessible from experiments. $\Delta\sigma_H(T,H)$ can be 
written out as:
\begin{eqnarray}
\Delta\sigma_{H}(T,H) & = & \Delta\sigma(T,H)-\Delta\sigma_{n}(T,H)\, \nonumber \\
& = & \Delta\sigma_{SF}(T,H)-\Delta\sigma_{SF}(T,0)
\label{Eq.correct-magneto}
\end{eqnarray}  
It has been very often assumed that the second term of the first equation can be neglected as being 
only weakly dependent on magnetic fields. However, our study clearly shows that this is not the case 
(see for instance the date displayed in fig.\ref{fig:MR-OPT}). Thus our method provides here a correct 
determination of $\Delta\sigma_H(T,H)$. 

\begin{figure}
\centerline{\includegraphics[width=0.8\columnwidth]{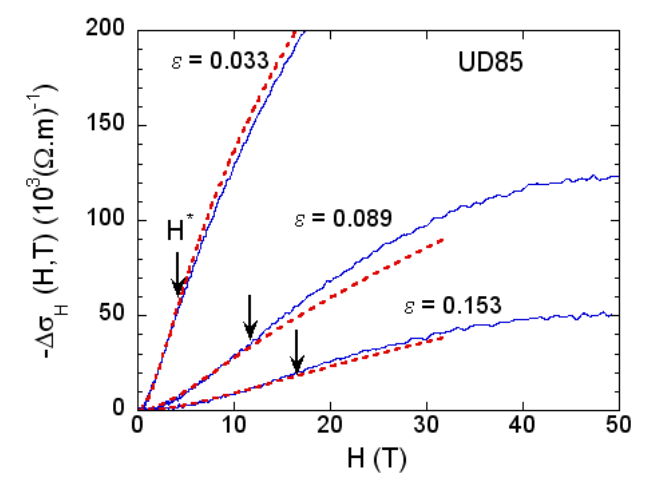}}
\caption{(color on line) Evolution of the fluctuation magnetoconductivity 
$-\Delta\sigma_{H}(T,H) =\Delta\sigma_{SF}(T,0)-\Delta\sigma_{SF}(T,H)$
as a function of $H$ for the UD85 sample at different temperatures: 87.5, 92.4 and 98.6K. 
The dotted lines represent the computed results from the ALO contribution with  $H_{c2}(0)=125(5)$T. 
They deviate from the data beyond the $H^{\star}$ field values shown by arrows \cite{FRA-PRB2011}.}
\label{Fig:fit-ALO}
\end{figure}

In the GL approach, the evolution of the fluctuation
magnetoconductivity with $H$ comes from the pair-breaking effect which leads to a $T_c$ suppression. 
In the case of interest, the major contribution results from the AL process, and more particularly 
from the interaction of the field with the carrier orbital (ALO) degrees of freedom. The detailed 
analysis and discussion of the fluctuation magnetoconductivity are reported in ref.\cite{FRA-PRB2011}. 
Fig.\ref{Fig:fit-ALO} shows the data for the UD85 sample together with the fits using the ALO 
expression with $H_{c2}(0)=125(5)$T being the only adjustable parameter. We can checked that, 
as predicted by the theory, the fits are valid as long as $H \lesssim H^{\star}(T)=\epsilon H_{c2}(0)$. 
This latter field defined above $T_c$ mirrors the upper critical field and has been called the 
"ghost critical field" by Kapitulnik et al \cite{Kapitulnik}. 

The different values of $H_{c2}(0)$ extracted from the low-field part of the magnetoconductivity data 
are reported in table\ref{tab:H}. They are surprisingly very close to those obtained for 
$H_{c}^{\prime}(0)$ in a completely different way. This gives strong weight to the consistency of our 
data analysis. 

The important result here is to show that the superconducting gap which is directly 
related to $H_{c2}(0)$ increases smoothly with increasing hole doping from the underdoped to the 
overdoped regime, contrary to the pseudogap which decreases. This is a strong indication that the gap determined here can thus be assimilated to the "small" gap detected recently by different techniques, while the pseudogap would be rather connected with the "large" gap \cite{Hufner}

\begin{table}[ht]
\caption{Values of $H_{c2}(0)$ extracted from the fluctuation magnetoconductivity. They are very close 
to the values of $H_{c}^{\prime}(0)$.}
\label{tab:H}     
\begin{tabular}{lllll}
\hline\noalign{\smallskip}
Sample & UD57 & UD85 & OPT93.6 & OD92.5  \\
\noalign{\smallskip}\hline\noalign{\smallskip}
$H_{c2}(0)$(T) & 90(10) & 125(5) & 180(10) & 200(10) \\
$H_{c}^{\prime}(0)$(T) & 86(10) & 115(5) & 155(10) & 207(10) \\
\noalign{\smallskip}\hline
\end{tabular}
\end{table}

\section{Influence of disorder}
\label{sec:disorder}
It is now well admitted that the properties of cuprates are strongly dependent on disorder. 
We have studied for long the effect of the introduction of controlled disorder by electron irradiation 
and the way it affects the transport properties \cite{FRA-PRL2003}. In particular, we have shown that similar
upturns of the low-T resistivity are found for controlled disorder in YBCO and in some "pure" low-$T_c$
cuprates, which indicates the existence of intrinsic disorder in those families \cite{FRA-EPL-MIT}. 

We have also carried out magnetoresistance measurements in some OPT93.6 and UD57 samples irradiated by
electrons. When $T_c$ is decreased by disorder, we find that both $T_{c}^{\prime}$ 
and $H_{c}^{\prime}(0)$ are also affected. The reduction in $T_c^{\prime}$ nearly follows that in $T_c$ 
for the underdoped sample while it is slightly larger for the OPT sample. Consequently, when $T_c$ is
decreased by disorder, the relative range of SCFs with respect to the value of $T_c$ expands considerably.
For instance, we still detect $T_{c}^{\prime} \sim 60$K in an UD57 irradiated sample with $T_c = 4.5$K. 

\begin{figure}[t]
\centerline{\includegraphics[width=0.8\columnwidth]{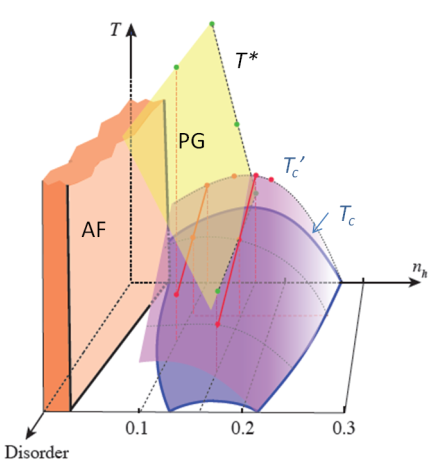}}
\caption{(color on line) Phase diagram constructed on the data points
obtained here, showing the evolution of $T_{c}^{\prime}$ the onset of
SCF, with doping and disorder. The fact that the pseudogap and the SCF
surfaces intersect each other near optimum doping in the clean limit is
apparent. These surfaces have been limited to experimental ranges where
they have been determined experimentally.}
\label{Fig:3Dphase-diagram}
\end{figure}

These results allow us to draw important conclusions on the cuprate phase diagram. Indeed, 
contrary to $T_c$, $T_c^{\prime}$ or $H_c^{\prime}$, the pseudogap temperature $T^{\star}$ has 
been found very early to be quite robust to disorder \cite{Alloul-PRL1991}. This is another indirect evidence that the 
pseudogap phase is not related to superconductivity. We want also to emphasize here that specific 
effects induced by disorder are probably at the origin of many confusions in the study of high-$T_c$
cuprates. This leads us to propose in fig.\ref{Fig:3Dphase-diagram} a 3D phase diagram where the 
effect of disorder has been introduced as a third axis. 
There, in the pure systems, the occurrence of SCFs and the difficulty to separate the SC gap from 
the pseudogap in zero-field experiments justifies that the $T_c^{\prime}$ line could often be taken as a 
continuation of the $T^{\star}$ line. 

It can also be seen in this figure that the respective evolutions with disorder 
of the SC dome and of the amplitude of the SCF range explains the phase diagram often shown in a 
low-$T_c$ cuprate such as Bi-2201 and sketched in fig.\ref{fig:phase-diagram}(a). Finally, for 
intermediate disorder, the enhanced fluctuation regime with respect to $T_c$ observed in the 
Nernst measurements for the La$_{2-x}$Sr$_x$CuO$_4$ can be reproduced as well \cite{Wang-PRB2006}.

\section{Conclusion}
\label{sec:conclusion}
We have presented here a thorough quantitative study of the SCFs which establishes that 
such data give important determinations of some thermodynamic properties of the SC state of 
high-$T_{c}$ cuprates. Those are not accessible otherwise, as flux flow dominates near $T_{c}$ in the 
vortex liquid phase and the highest fields available so far are not sufficient to reach the normal state 
at $T=0$. It has allowed us to demonstrate that the pairing energy and SC gap both increase 
with doping, confirming then that the pseudogap has to be assigned to an independent magnetic order 
or crossover due to the magnetic correlations.

\begin{acknowledgement}
We thank D. Colson and A. Forget for providing the single crystals used in this work.
We acknowledge collaboration with C. Proust, B. Vignolle and G. Rikken at the LNCMI.
This work has been performed within the "Triangle de la Physique" and was supported by 
ANR grant "OXYFONDA" NT05-4 41913. The experiments at LNCMI-Toulouse were funded by the FP7 I3 EuroMagNET. 
\end{acknowledgement}

\end{document}